\documentclass[reprint, aps, prl, nofootinbib]{revtex4-2}  %

\usepackage{graphicx}%
\usepackage{dcolumn}%
\usepackage{bm}%
\usepackage[linktocpage,colorlinks,pdfusetitle,allcolors=blue]{hyperref}
\usepackage{amsmath, amssymb}
\usepackage{siunitx}

\DeclareSIUnit[]\admmass{\text{\ensuremath{M}}}

\begin{document}

\title{Does charge matter in high-energy collisions of black holes?}

\author{Gabriele Bozzola}
\email{gabrielebozzola@email.arizona.edu}
\affiliation{Department of Astronomy, University of Arizona, Tucson, AZ, 85721, USA}

\date{\today}

\begin{abstract}
  We perform numerical-relativity simulations of high-energy head-on collisions
  of charged black holes with the same charge-to-mass ratio $\lambda$. We find that
  electromagnetic interactions have subdominant effects already at low Lorentz
  factors $\gamma$, supporting the conjecture that the details of the properties of
  black holes (e.g., their spin or charge) play a secondary role in these
  phenomena. Using this result and conservation of energy, we argue these events
  cannot violate cosmic censorship.
\end{abstract}

\maketitle

\paragraph{\textbf{Introduction}}

High-energy collisions of black holes are excellent laboratories to probe
general relativity and to study the theory under extreme conditions (for
reviews, see relevant sections in~\cite{Sperhake2014, Choptuik2015}). Due to its
highly dynamical nature, the problem is best approached with numerical
calculations, as the ones that opened this line of research in
2008~\cite{Sperhake2008,Shibata2008}. Since then, studies have explored most of
the possible variables (mass, impact parameter,
spin~\cite{Sperhake2008,Shibata2008,Sperhake2009,Healy2016,Sperhake2011,
  Sperhake2011b,Gold2014c, Sperhake2016, Sperhake2019, Andrade2019,
  Andrade2020}), with the noticeable exclusion of charge (electric, or
associated to a generic $\text{U(1)}$ %
field). In this Letter, we tackle this long overdue problem and present
general-relativistic simulations of head-on collisions of black holes with the
same charge.

One of our main objectives is to test whether ``matter matters'' in the
ultra-relativistic regime~\cite{tHooft1987,Amati1987,Banks1999}. According to
this idea, the details of the properties of bodies (e.g., their spin, or
composition) are irrelevant in collisions with high center-of-mass energy. The
conjecture originates from considering that ultra-relativistic mergers are
dominated by the kinetic energy, so the details of the interaction are
unimportant. This has been verified numerically for
spinning~\cite{Sperhake2011b, Sperhake2013} and non-spinning black
holes~\cite{Sperhake2008, Shibata2008, Sperhake2011}, as well as for boson
fields~\cite{Choptuik2010}, perfect fluids~\cite{East2013, Rezzolla2013}, and
plane waves~\cite{Pretorius2018}. The problem has also been studied in higher
dimensions~\cite{Witek2010, Okawa2011, Sperhake2019, Andrade2019, Andrade2020},
where, in case of AdS$_{5}$, it is relevant for gauge-gravity dualities. This
conjecture is also at the basis of Monte Carlo event
generators~\cite{Cavaglia2007,Dai2008,Frost2009} for microscopic black holes in
particle accelerator. Here, we test this hypothesis for black holes with charge,
parameter that has not been considered so far.

Our second goal is to check if it is possible to form naked singularities with
ultra-relativistic collisions, verifying whether the cosmic censorship
conjecture holds. Testing this has been a recurring theme in this line of
research (e.g.~\cite{Sperhake2008, Sperhake2009, Okawa2011, Andrade2020}), but
no violation has been found so far in four-dimensional spacetimes. High-energy
collisions of charged black holes are a particularly interesting setting to
investigate this idea because charge is another way, together with spin, to
reach black-hole extremality. Kerr-Newman spacetimes with too much charge and/or
spin compared to their mass do not have horizons~\cite{Wald1984}, so
overcharging or overspinning a black hole would be a way to form a naked
singularity. Because of the emission of energy, ultra-relativistic collisions
might lead to conditions in which the remnant would be ``over-extremal'', and
create a naked singularity. In the case of spinning black holes, this is avoided
by radiating away the excess angular momentum. However, charge is conserved and
cannot be radiated away, constituting a significant difference compared to spin.
Moreover, if charge does not matter, the colliding black holes will always merge
and will not repel due to electrostatic interaction. So, if the formation of
naked singularities is avoided, it is interesting to understand how this is
achieved.

This Letter focuses on testing whether charge is important in the context of
high-energy collisions and whether naked singularities can form in this
environment. Our goal is not to perform a high-precision study, which would
require extreme numerical resolution and sophisticated initial data (see,
e.g.~\cite{Ruchlin:2014zva, Healy2016}), but we aim to describe the general
features of the phenomenon. Our main conclusion is that we find evidence that,
even at low value of the boost factor $\gamma$, important gauge-independent
quantities do not depend on the charge, supporting the idea that \emph{charge
  does not matter} in ultra-relativistic collisions. Having found no evidence
that all the kinetic energy in the system can be radiated away, we argue that
ultra-relativistic collisions of black holes with the same charge do not form
naked singularities. These conclusions are robust despite the overall accuracy
of our simulations of order \SI{10}{\percent}. In general, our full general
relativistic calculations show that the problem can be well-understood with
simple semi-classical arguments, which we present below.

The Letter is structured as follows. First, we describe our theoretical and
numerical setup. Then, we report the results and our interpretation, and
finally, we give some concluding remarks. We use Gaussian units with
$G = c = 4 \pi \epsilon_{0} = 1$, and we report results in terms of
$\si{\admmass} = M_{1} + M_{2}$, where $M_{1}$ and $M_{2}$ are the individual
Christodoulou masses~\cite{Christodoulou1971, Bozzola2019}.

\paragraph{\textbf{Setup}} We solve the Einstein-Maxwell equations in the $3+1$
decomposition of the spacetime~\cite{Arnowitt:1962hi,Arnowitt2008} (see
also~\cite{Alcubierre:2008it,Baumgarte:2010nu,Shibata2016b}) for head-on
collision of equal-mass, equal-charge black holes with charge-to-mass-ratio
$\lambda \in \{0, 0.2, 0.4, 0.6, 0.8\}$ and initial linear momentum
$P\slash \si{\admmass} \in \{0.2, 0.4, 0.6\}$. We use the \texttt{Einstein
  Toolkit}~\cite{Loffler:2011ay,EinsteinToolkit:2021_05} for the numerical
integration and \texttt{kuibit}~\cite{kuibit} for the analysis. We adopt the
same setup as in~\cite{Bozzola2021}, where we provide a more in-depth
discussion. Note that, with the exception of \texttt{kuibit} and
\texttt{TwoChargedPunctures} (see below), we use the same computational tools
that have been extensively employed in this line of
research~(e.g.~\cite{Sperhake2008, Sperhake2009, Sperhake2011, Sperhake2011b,
  Zilhao2012, Sperhake2013, Zilhao2013, Healy2016, Sperhake2016, Sperhake2019}).

We generate constraint-satisfying initial data with
\texttt{TwoChargedPunctures}~\cite{Bozzola2019} for two black holes with
masses~\cite{Christodoulou1971, Bozzola2019}
$M_{1} = M_{2} = \SI{0.5}{\admmass} $ and charge-to-mass ratio
$\lambda_{1} = \lambda_{2} = \lambda$. The two punctures are aligned along the $z$ axis with an
initial separation of $\SI{150}{\admmass}$. In the limit of infinite separation
of in the case of isolated black holes,
\texttt{TwoChargedPunctures}~\cite{Bozzola2019} reduced to
Reissner-Nordstr{\"o}m in isotropic coordinates. The boost factor is controlled
by the Bowen-York momentum $P$, an input parameter in
\texttt{TwoChargedPunctures}, which is equal to the Arnowitt-Deser-Misner
(ADM)~\cite{Arnowitt:1962hi,Arnowitt2008} linear momentum for a case of a single
black hole~\cite{Bozzola2019} and corresponds to Lorentz factor of
$\gamma = \sqrt{1 + 4 P^{2} \slash M^{2}}$. \texttt{TwoChargedPunctures} employs the
conformal-traceless-traverse approach~\cite{York1971, Bowen1985, Alcubierre2009,
  Bozzola2019}, extending what is done by the well-known
\texttt{TwoPunctures}~\cite{Ansorg2004} pseudo-spectral solver for the uncharged
case. In particular, the code assumes conformal flatness and
Reissner-Nordstr{\"o}m electromagnetic fields. This leads to ``junk'' radiation,
especially in the electromagnetic sector, that can be up to a few percent of the
total energy. The initial separation is large enough that we can isolate the
real signal from the spurious one (see also Supplemental Material).

We evolve the spacetime and electromagnetic fields with the \texttt{Lean} and
\texttt{ProcaEvolve} codes~\cite{canudacode, canuda, Sperhake2007,Zilhao2015}.
\texttt{Lean} implements the Baumgarte-Shapiro-Shibata-Nakamura (BSSN)
formulation of Einstein's equation~\cite{Shibata1995,Baumgarte1998} and the
moving puncture approach, while \texttt{ProcaEvolve} evolves the electromagnetic
vector potential to maintain the magnetic field divergenceless and has a
constraint-damping scheme for the Gauss constraint. We use the Lorenz gauge for
the electromagnetic potential, the $1 + \log$ and $\Gamma$-freezing slicing
conditions for the lapse function and shift
vector~\cite{Alcubierre:2002kk,van-Meter2006,Hinder:2013oqa}.

The simulations are on \texttt{Carpet}~\cite{Schnetter:2003rb} Cartesian grids
with octant symmetry, with two centers of refinement (one tracking the puncture,
and the other fixed in the center) and 13 levels. The outer boundary is placed
at least at \SI{600}{\admmass}, where it is not in causal contact with the inner
part of the grid throughout the duration of the simulations. We use the
continuous Kreiss-Oliger dissipation introduced in~\cite{Bozzola2021}.

Since the size of the horizons depends on the charge-to-mass ratio $\lambda$, we
change the resolution to ensure that the black holes are always resolved with at
least 80 points. We estimate the initial horizon radius as if it was a
Reissner-Nordstr{\"o}m black hole in isotropic coordinates~\cite{Bozzola2019}
and set the finest grid spacing to
$\Delta x_{\text{finest}} = \sqrt{1 - \lambda^{2}}\slash 320\,\si{\admmass}$.\footnote{In
  isotropic coordinates, the horizon radius for a Reissner-Nordstr{\"o}m black
  hole with mass $M_{1} = \SI{0.5}{\admmass}$ and charge $Q_{1} = \lambda M_{1}$ is
  $\sqrt{1 - \lambda^{2}} \slash 4\,\si{\admmass}$.} Depending on the charge, this can lead
to resolutions up to $\si{\admmass}\slash 550$.

We find that our simple prescription for the grid resolution is effective in
properly resolving the black holes. We locate apparent horizons with
\texttt{AHFinderDirect}~\cite{Thornburg:2003sf}, and compute their properties
with \texttt{QuasiLocalMeasuresEM}~\cite{Bozzola2019}, an extension of
\texttt{QuasiLocalMeasures}~\cite{Dreyer2003} for full Einstein-Maxwell
theory~\cite{Ashtekar2000,Ashtekar2001,Ashtekar2004}. At the level of the
initial data, we find that the horizons are coordinate ellipsoids covered by at
least 40 points along the semi-minor axis. Then, the horizons expand and for
most of the simulation our grid resolves the semi-minor axis with at least 120
points. The merger remnant is resolved even better. As a result, the quasi-local
properties are well-behaved in all our simulations (e.g., charge is conserved at
better than \SI{0.6}{\percent}).

We extract radiation with the Newman-Penrose formalism~\cite{Newman1962, Witek2010,
  Zilhao2015} at finite extraction radii ranging from $\SI{80}{\admmass}$ to
$\SI{200}{\admmass}$. We note that, while the properties of the horizons are
remarkably stable, interpolation across several refinement boundaries and the
truncation error in the wave zone lead to noisy electromagnetic waves (see
Supplemental Material).

\paragraph{\textbf{Results}}

The main conclusion from our simulations of high-energy head-on collisions of
black holes is that charge does not matter for a number of gauge-independent
quantities. Before we present our results in detail, we define quantitatively
what we mean by ``charge does not matter''. In Newtonian physics, the problem of
two charged point masses is mathematically equivalent to the purely
gravitational one upon rescaling of $G$ by a factor $(1-\lambda^{2})$. This simple
scaling is surprisingly effective in predicting results of fully
general-relativistic
calculations~\cite{Zilhao2012,Zilhao2013,Bozzola2020,Bozzola2021}. Therefore, if
charge mattered, we would expect most results (e.g., amplitude of $\Psi_{4}$) to
vary with factors $(1-\lambda^{2})$ for varying $\lambda$ and fixed $P$. Conversely, if
charge did not matter, all the results should become approximately the same
within our error (see, Supplemental Material).

\begin{figure}[htbp]
  \centering
  \includegraphics[width=\linewidth]{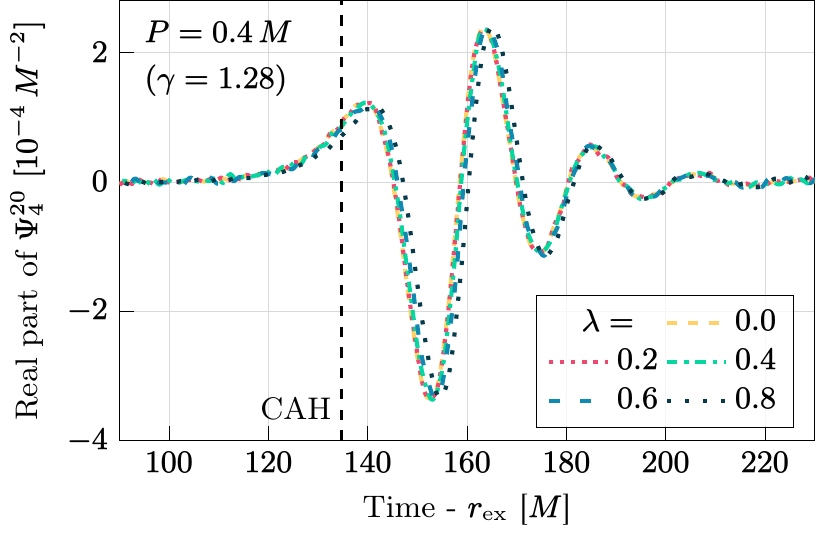}
  \caption{Real part of the dominant multipolar component of the Newman-Penrose
    scalar $\Psi_{4}$ ($l=2$, $m=0$) for simulations with different charge-to-mass
    ratio $\lambda$ as extracted at radius $r_{\text{ex}} = \SI{131.430}{\admmass}$.
    Note that no time-shift was applied to the signals: the almost identical
    alignment indicates that charge has a negligible influence in these
    collisions. The time of formation of the horizon is also nearly insensitive
    to the value of charge. }%
  \label{fig:psi4-real}
\end{figure}

\begin{figure}[htbp]
  \centering
  \includegraphics[width=\linewidth]{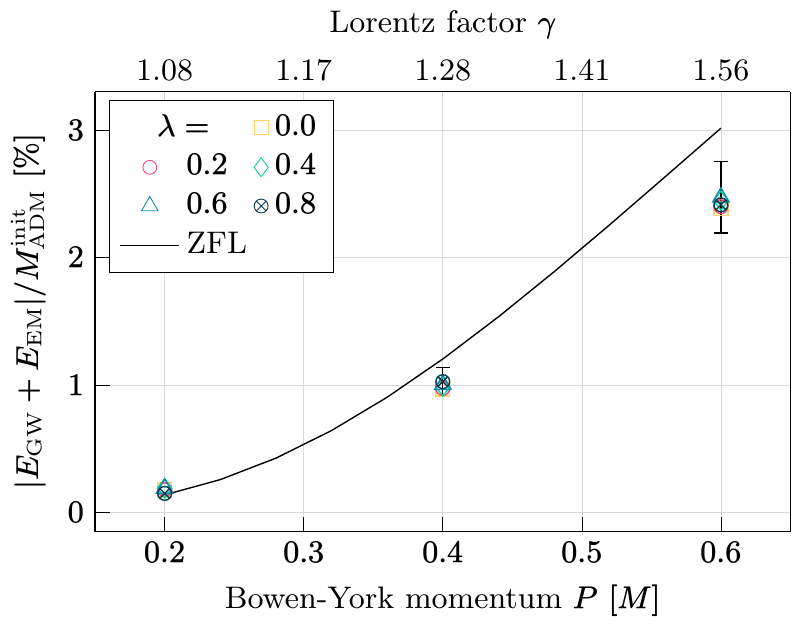}
  \caption{Total energy lost by gravitational and electromagnetic waves
    normalized to the initial ADM mass. At any given Bowen-York momentum $P$,
    the energy lost for different values of $\lambda$ is the same (within our error,
    see Supplemental Material). The black line is the Zero-Frequency-Limit (ZFL)
    prediction~\cite{Smarr1977} (Eq.~\eqref{eq:zfl} with $E_{\infty} = 0.13$), which
    has been shown to be accurate for uncharged collisions~\cite{Sperhake2008,
      Healy2016}.}%
  \label{fig:energy}
\end{figure}

We demonstrate that charge has negligible influence in the dynamics of
high-speed mergers by discussing some key properties of the gravitational waves
and of the horizons. In Fig.~\ref{fig:psi4-real}, we present the real part of
the dominant mode of the Newman-Penrose scalar $\Psi_{4}$ ($l=2$, $m=0$) for
simulations with fixed Bowen-York momentum $P = \SI{0.4}{\admmass}$ and varying
charge-to-mass ratios $\lambda$. We do not apply any time-shift or any other
transformation to align the signals. The good alignment indicates that charge
does not have a strong impact in the event (compare this with Fig.~5
in~\cite{Zilhao2012}, where signals had to be scaled by the factor $1-\lambda^{2}$;
see also Supplemental Material for a more detail comparison). We find the same
properties as in the uncharged case~\cite{Sperhake2008}: there is a precursor
signal, a main burst after the formation of the apparent horizon, and the
ringdown. The time of formation of the common apparent horizon (vertical dashed
line) is nearly independent of the charge and the peak of the signal occurs
always approximately $\SI{15}{\admmass}$ after this time.
The total energy lost by gravitational and electromagnetic waves is reported in
Fig.~\ref{fig:energy}. Collisions from zero initial momenta were studied
in~\cite{Zilhao2012}, where it was found that there is a significant
contribution from electromagnetic fields to the energy radiated (up to
\SI{25}{\percent}). Instead, we never find large amounts of electromagnetic waves in
our simulations, and almost all the energy is lost through gravitational waves.
Fig.~\ref{fig:energy} shows how all our simulations at fixed $P$ radiate the
same amount of energy irrespective of $\lambda$ (within our error, see Supplemental
Material). In Fig.~\ref{fig:energy}, we also plot the estimate of the energy
lost in ultra-relativistic collisions obtained in the Zero-Frequency-Limit
(ZFL)~\cite{Smarr1977}, which has been shown to be a good approximation to the
fractional energy lost ${E_{\text{GW}}}\slash{M_{\text{ADM}}}$ for collisions in
absence of charge~\cite{Sperhake2008, Healy2016}. According to this formalism,
${E_{\text{GW}}}\slash{M_{\text{ADM}}}$ scales as:
\begin{equation}
  \label{eq:zfl}
  \frac{E_{\text{GW}}}{M_{\text{ADM}}} = E_{\infty} \left(\frac{1 + 2 \gamma^{2}}{2 \gamma^{2}} + \frac{(1 - 4\gamma^{2}) \ln (\gamma + \sqrt{\gamma^{2} - 1})}{2 \gamma^{3} \sqrt{\gamma^{2} - 1}}  \right)\,,
\end{equation}
where $E_{\infty}$ is the energy lost for infinitely boosted black holes, which has
numerically been calibrated to be approximately \num{0.13}. Our simulations also
find a good level of agreement with the ZFL estimate.

Finally, we consider the remnant properties. We find that the fractional
difference of the quasi-local mass of the final black hole between the charged
and uncharged cases is always below \SI{1}{\percent}. This implies that the mass of the
remnant does not depend on $\lambda$ at the level of our accuracy (see Supplemental
Material). Note, however, that apparent horizons are not completely gauge
invariant as they depend on the spacetime slicing.

Our simulations demonstrate that even with small boosts ($\gamma \approx 1.1$) \emph{charge
  does not matter} in the dynamics of the event and in a number of
gauge-independent quantities, or if it did, it would do so only at the percent
level (contrarily to what happens for $\gamma=1$~\cite{Zilhao2012}). We can build
intuition on why this happens with the following qualitative semi-classical
argument. Consider a head-on collision of two black holes with mass $\mathcal{M}$, charge
$\mathcal{Q} = \lambda \mathcal{M}$, Lorentz factor $\gamma$, and infinite initial distance. Initially, the
interaction is negligible and the motion is completely determined by the initial
velocity. The separation $d_{\text{EM}}$ at which the electromagnetic
interaction starts to be important is when the magnitude of its associated
energy ($\lambda^{2} \mathcal{M}^{2} \slash d_{\text{EM}}$) is comparable to the kinetic energy
($2 (\gamma - 1) \mathcal{M}$):\footnote{The gravitational interaction starts to be important
  at larger separations. However, this increases the kinetic energy and only
  makes the conclusions stronger. }
\begin{equation}
  \label{eq:sep-charge-matter}
  d_{\text{EM}} = \frac{\lambda^{2} \mathcal{M}}{2(\gamma - 1)} = \frac{\lambda^{2}}{4} \frac{\mathcal{M}_{\text{ADM}}}{\gamma(\gamma-1)}\,,
\end{equation}
where we used that $\mathcal{M}_{\text{ADM}} = 2 \gamma \mathcal{M}$. For separations that are much
larger than this value, the bodies can be considered non-interacting, so charge
does not matter. In classical physics, particles will always reach
$d_{\text{EM}}$, where they start to be repelled by the electrostatic force.
This is not what happens for black holes, where there is another length-scale
that we need to consider and that drastically alters this picture. Assuming that
all the initial energy ends up in the remnant, and calling
$\mathcal{R} = 2 \mathcal{M}_{\text{ADM}}$ its Schwarzschild radius, we expect $\mathcal{R}$ to be where
general-relativistic effects to be dominant (consider, for example, the hoop
conjecture~\cite{Thorne1972}). When the two initial horizons get closer than
$\mathcal{R}$, they stick together as a newly formed remnant, overcoming the electrostatic
repulsion. So, if $d_{\text{EM}} \ll \mathcal{R}$, electromagnetism starts to be dominant
only after the formation of a common apparent horizon and charge would be
unimportant. We conclude that charge does not matter when
$d_{\text{EM}} \slash \mathcal{R} \ll 1$, and, according to our simple model,
${d_{\text{EM}}}\slash{\mathcal{R}} = {\lambda^{2}}\slash{[8\gamma(\gamma-1)]}$. This value is smaller or much
smaller than 1 for all $\lambda$ and $P$ we considered, consistently with the results
of our numerical-relativity simulations.

\begin{figure}[htbp]
  \centering
  \includegraphics[width=\linewidth]{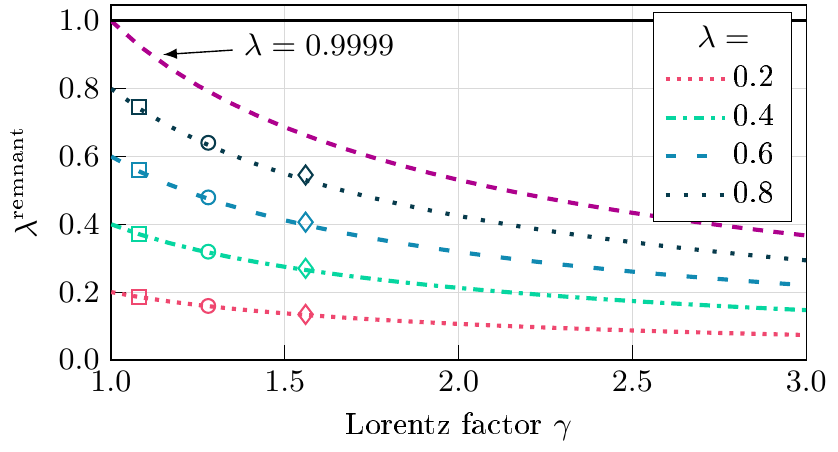}
  \caption{Charge-to-mass ratio $\lambda^{\text{remnant}}$ for the remnant left by a
    merger of two equal-mass black holes with initial Lorentz factor $\gamma$ and
    charge-to-mass ratio $\lambda$. The curves are obtained with
    Eq.~\eqref{eq:lambda_rem} assuming $\Upsilon(\gamma) = 0$ (expected from the fact that
    charge does not matter in the energy emitted in these mergers) and the
    markers are the values from our simulations. The figure seems to hint that
    the only case where we can obtain an overcharged remnant is with $\lambda,\gamma\to1$,
    where our approximations break down and previous studies found no
    violation~\cite{Zilhao2012}.}%
  \label{fig:cosmic}
\end{figure}

Established that charge plays a subdominant role in the dynamics of the event
under consideration, we can now turn to the problem of cosmic censorship. We
argue that the conjecture is not violated in ultra-relativistic head-on
collisions of charged black holes on the grounds that the final black hole
always has $\lambda^{\text{remnant}} < 1$ for any given initial charge and momentum.
We tackle this problem with conservation arguments. Consider two black holes
with Christodoulou mass $\mathcal{M}$, charge $\mathcal{Q} = \lambda \mathcal{M}$ boosted with Lorentz factor $\gamma$
and initial separation such that they can be considered non-interacting.
Conservation of energy implies that the mass of the remnant has to be
$\mathcal{M}^{\text{remnant}} = \mathcal{M}_{\text{ADM}} - E_{\text{GW}} - E_{\text{EM}}$, where
$E_{\text{GW}}$ and $E_{\text{EM}}$ are the energies carried away by
gravitational and electromagnetic waves respectively. Let us define
$\Upsilon(\gamma) = {E_{\text{EM}}}\slash{E_{\text{GW}}}$ and
$Z(\gamma) = {E_{\text{GW}}}\slash{ \mathcal{M}_{\text{ADM}}}$. As shown in Fig.~\ref{fig:energy},
the ZFL approach provides a good approximation to $Z(\gamma)$, so we can use the
expression in Eq.~(3) in~\cite{Sperhake2008}, noting that $Z(\gamma) \lesssim 0.14$ for any
value of $\gamma$~\cite{Sperhake2008, Healy2016}. Conversely, we do not have a good
formula for $\Upsilon(\gamma)$. In~\cite{Zilhao2012} it was found that $\Upsilon(1) \approx \lambda^{2} \slash 4$,
and our simulations show that $\Upsilon(\gamma) \ll \lambda^{2} \slash 4$ even for low values of $\gamma$, in
accordance with the conjecture that charge does not matter. So, assuming that
the conjecture is true, $\Upsilon(\gamma)$ has to be at least bound. Dividing the equation
of energy conservation by $\mathcal{M}_{\text{ADM}}$ and using
$\mathcal{M}^{\text{remnant}} = 2 \mathcal{Q} \slash \lambda^{\text{remnant}}$ (charge is conserved) and
$\mathcal{M}_{\text{ADM}} = 2 \gamma \mathcal{Q} \slash \lambda$, we find that
\begin{equation}
  \label{eq:lambda_rem}
  \lambda^{\text{remnant}}(\gamma) =  \left[ \frac{1}{1 -  \left( 1 + \Upsilon(\gamma) \right) Z(\gamma)} \right] \frac{\lambda}{\gamma} \,.
\end{equation}
Given that $\Upsilon(\gamma)$ and $Z(\gamma)$ are bound, there exists a constant $C$ such that
the term in the brackets is smaller than $C$ for all $\gamma$. Hence,
$\lambda^{\text{remnant}} \le C \lambda \slash \gamma$, indicating that $\lambda^{\text{remnant}}$ decreases
with $\gamma$. In Fig.~\ref{fig:cosmic}, we show Eq.~\eqref{eq:lambda_rem} by
reporting the values of $\lambda^{\text{remnant}}$ predicted for various $\lambda$ assuming
$\Upsilon(\gamma) \ll 1$. We overlay the result of our simulations with markers, which are in
excellent agreement. Since in the limit of $\gamma\to\infty$, Eq.~\eqref{eq:lambda_rem}
predicts that $\lambda^{\text{remnant}}$ goes to zero, we find agreement with the
conjecture that matter does not matter and we conclude ultra-relativistic
head-on collisions of charged black holes should not be expected to form naked
singularities. This result is robust and only depends on the assumption that
electromagnetic waves cannot radiate away all the additional kinetic energy, as
our general relativistic calculations show.

\paragraph{\textbf{Conclusions}}

Ultra-relativistic collisions of black holes are fertile ground for theoretical
studies in general relativity and high energy physics. In this Letter, we
presented the first results on high-energy head-on mergers of charged black
holes. We found that the intuition built with simple semi-classical arguments
carries over to full general relativity. First, we found that charge does not
play an important role, supporting the conjecture that \emph{matter does not
  matter}. This is an important step in claiming that the conclusion holds for
generic four-dimensional general-relativistic black holes. This result is also
important in the context of the production of microscopic black holes in
particle accelerators and cosmic rays. We also argued that, as a result, we
should not expect the formation of naked singularities in this kind of event.

Given that the expectation that charge is unimportant is met even with
relatively low boosts, we anticipate that varying the other variables that were
not considered here (mass, impact parameter, charge, spin) will yield the same
results as the uncharged case. This should be tested, along with expanding the
current study to more extreme $\lambda$ and $P$ and increasing the accuracy. This
might require enhancement in the initial data (e.g., by using better guesses for
the electromagnetic fields and by lifting the assumption of conformal flatness)
and a reduction in the error budget (e.g., by reducing initial data ambiguity,
increasing the accuracy in the wave zone--possibly with multi-patch
grids~\cite{Pollney:2009yz}--and performing interpolation of waves to infinity).

\begin{acknowledgments}
  G.\ B.\ is indebted to Vasilis Paschalidis for several insightful
  conversations and comments on the manuscript. This research was made possible
  by the developers and maintainers of the open-source codes that we used.
  \texttt{kuibit}~\cite{kuibit} uses \texttt{NumPy}~\cite{NumPy},
  \texttt{SciPy}~\cite{SciPy}, and \texttt{h5py}~\cite{h5py}. This work was
  supported by NSF Grant PHY-1912619 to the University of Arizona, a Frontera
  Fellowship by the Texas Advanced Computing Center (TACC), and NASA Grant
  80NSSC20K1542. Frontera~\cite{Frontera2020} is founded by NSF grant
  OAC-1818253. Computational resources were provided by the Extreme Science and
  Engineering Discovery Environment (XSEDE) under grant number TG-PHY190020.
  XSEDE is supported by the NSF grant No.\ ACI-1548562. Simulations were
  performed on \texttt{Stampede2}, which is funded by the NSF through award
  ACI-1540931.
\end{acknowledgments}

\section*{Supplemental Material: Comparison with previous studies}%
\label{sec:append-comp-prev}

\begin{figure}[htbp]
  \centering
  \includegraphics[width=\linewidth]{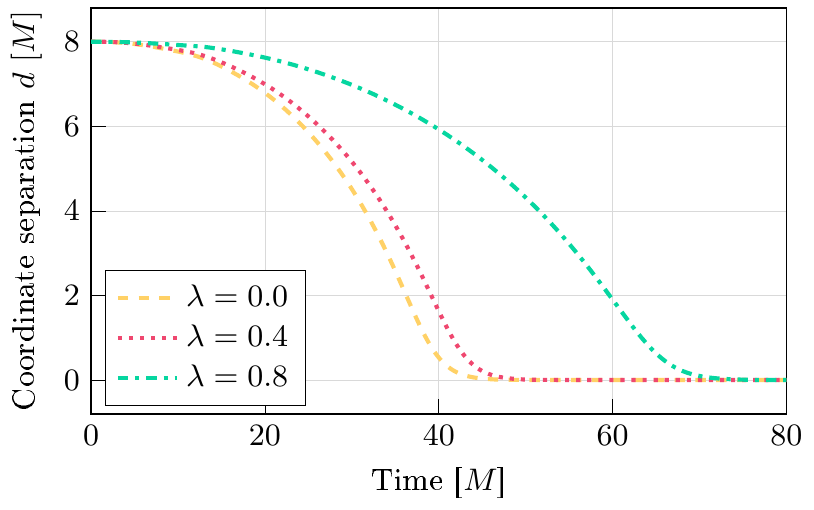}
  \caption{Coordinate separation as a function of time for mergers of
    equal-mass, equal-charge black holes with initial separation of
    \SI{8}{\admmass} zero boost. The plot shows that the dynamics of the event
    depend on the charge-to-mass ratio $\lambda$, in contrast to what found for the
    case with larger initial boost (see, Fig.~\ref{fig:psi4-real}).}%
  \label{fig:sep_8M}
\end{figure}

\begin{figure}[htbp]
  \centering
  \includegraphics[width=\linewidth]{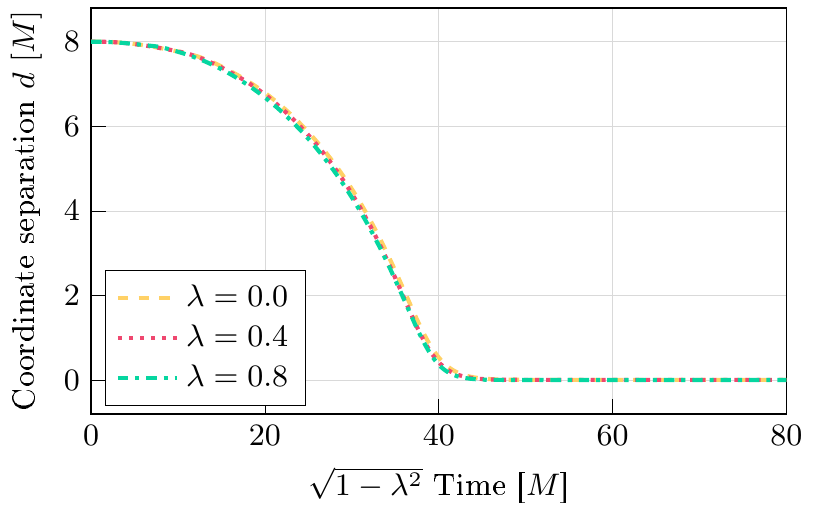}
  \caption{Same as Fig.~\ref{fig:sep_8M}, but with time rescaled by
    $\sqrt{1-\lambda^{2}}$. This is the same result obtained in~\cite{Zilhao2012},
    where it was found that charge matters for mergers with zero initial
    velocity.}%
  \label{fig:sep_8M_scaled}
\end{figure}

Our work extends previous studies~\cite{Zilhao2012} to the case with non-zero
initial boost. In~\cite{Zilhao2012}, it was found that charge plays an important
role in the merger, while we find that at $\gamma \approx 1.1$ the effects are negligible.
Our set of simulations does not capture the transition between the two regimes.
To do so one would need to perform simulations with smaller $\gamma$, calculations
that are computationally expensive with a setup like ours. Our simulations are
at high resolution (up to $\si{\admmass}\slash 550$ on the finest level) and with an
initial separation of \SI{150}{\admmass}, leading to a long time to merger. To
further connect our results with the ones in~\cite{Zilhao2012}, we reproduced
selected simulations in~\cite{Zilhao2012} (where the initial separation and
resolution are smaller than the ones used in this work). Fig.~\ref{fig:sep_8M}
shows the coordinate separation for mergers of charged black holes with no
initial boost. While this is not a gauge-invariant quantity, it is suggestive of
what happens when $\gamma=0$: systems with different values of charge behave
differently. More specifically, Fig.~\ref{fig:sep_8M_scaled} shows that this
behavior is fully captured by the classical arguments that predict a scaling as
$\sqrt{1 - \lambda^{2}}$. This is the main result obtained in~\cite{Zilhao2012}. These
two plots are in contrast with Fig.~\ref{fig:psi4-real}, where no scaling was
needed to obtain overlap between the different signals.

\section*{Supplemental Material: Error estimate}%
\label{sec:append-error-budg}

In our study, we have both numerical and systematic sources of error. Here, we
discuss them and provide an estimate of the overall accuracy of our simulations.
Note that we do not need an accurate estimate of the error, since the results we
presented are not affected by this.

\begin{figure}[htbp]
  \centering
  \includegraphics[width=\linewidth]{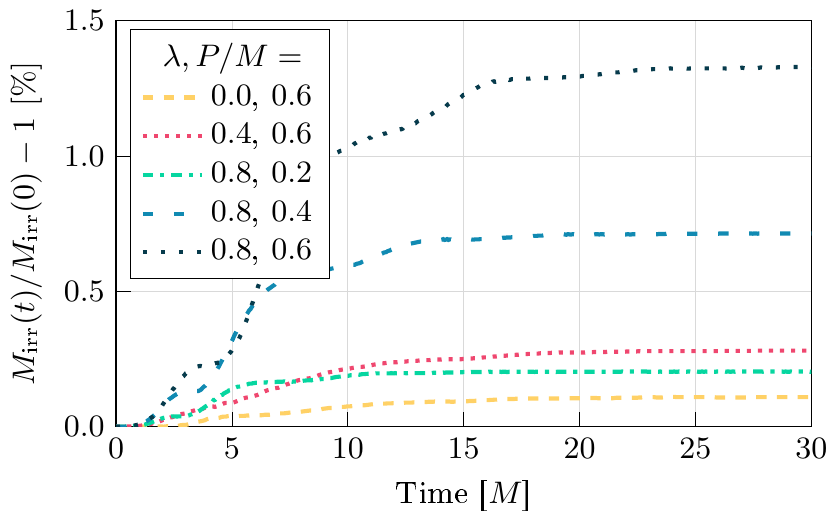}
  \caption{Variation in the irreducible mass of one of the horizons as measured
    by \texttt{QuasiLocalMeasuresEM} in the first \SI{30}{\admmass} of
    evolution. We use this quantity to asses the quality of the initial data and
    the relaxation time. Our assumption on conformal flatness and the choice of
    initial electromagnetic fields limits the maximum linear momentum that can
    be generated. Higher charge-to-mass ratio $\lambda$ and momentum $P$ lead to
    initial data further from equilibrium. The masses does not grow past the
    time plotted here, so we can see that the relaxation time is of order of
    \SI{10}{\admmass}.}%
  \label{fig:irr_mass}
\end{figure}

The results of this Letter hinge upon comparing simulations with same value of
$P$ and different value of $\lambda$. This comparison is not straightforward for
several reasons. First, the assumptions that go in the initial data (mainly the
conformal flatness and the initial electromagnetic fields) lead to initial black
holes that are not perfectly in equilibrium. Hence, the initial data relaxes to
a new configuration with different values of mass and momentum.\footnote{One
  could start the simulation, let the initial data relax, and measure the new
  mass and momentum. However, due to the lack of a well-defined notion of
  quasi-local linear momentum~\cite{Krishnan2002,Ashtekar2004}, this task is not
  trivial and beyond the goals of this Letter.} We report how the irreducible
mass evolves for a five representative cases in Fig.~\ref{fig:irr_mass}
For the simulations explored in this Letter, the change is always below
\SI{1.5}{\percent}, providing an estimate of the accuracy of the initial mass
and momentum. For a fixed momentum $P$, simulations behave differently depending
on $\lambda$, with higher charges starting further from equilibrium. The assumptions
in the initial data also limit the maximum boost attainable, as for larger
values of $P$, the initial data relaxes to a new configuration with a smaller
momentum. Second, in our simulations, we fix the coordinate distance between the
two punctures, which, by definition, is a gauge-dependent quantity. Simulations
have different coordinates, so it is not possible to compare them directly.
In~\cite{Zilhao2012}, a more robust way to compare different simulations of
charged black holes is used and the proper distance between the horizons is
computed. It was found (Table I in~\cite{Zilhao2012}) that the difference for
different values of charge is at the level of the percent. While the initial
data are different, we expect similar results to hold here. Moreover, since the
black holes start from a distance of \SI{150}{\admmass}, they are almost
isolated (with error at the level of the percent). Third, initial data contains
junk radiation, especially electromagnetic (due to the choice of initial
electromagnetic fields, which is not well adapted to cases with high boost). For
large charge and boosts, this radiation can be up to a few percent of the total
energy in the spacetime and is another source of error and limit to our study.
Because of the different amount of junk radiation and the interaction energy,
simulations will have slightly different ADM mass for a fixed value of $P$, with
a maximum variation of \SI{2}{\percent}. This directly affects the denominator in
Fig.~\ref{fig:energy}. These three effects introduce a fundamental systematic
error of order of percent in our work that would still be present in the limit
of infinite numerical resolution.

\begin{figure}[htbp]
  \centering
  \includegraphics[width=\linewidth]{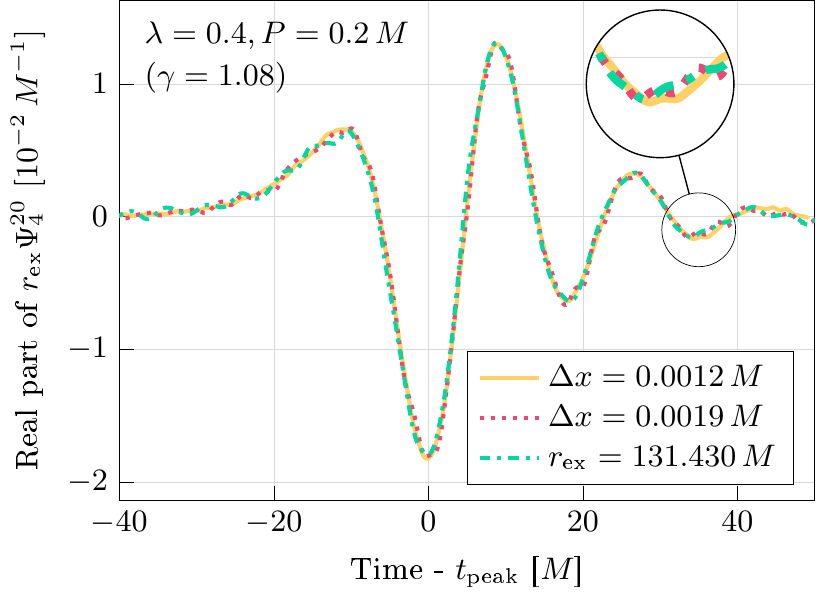}
  \caption{Comparison between the $l=2, m=0$ mode of the Newman-Penrose scalar
    $\Psi_{4}$ for simulations with different resolutions and extracted at
    different radii $r_{\text{ex}}$ ($r_{\text{ex}} = \SI{97.143}{\admmass}$).
    The plot shows good convergence, up to a well-known high-frequency
    noise~\cite{Zlochower2012} (as shown in the inset). We obtain an error
    estimate comparing the total energy radiated away for simulations with
    different extraction radii and/or resolution. }%
  \label{fig:psi4_conv}
\end{figure}

The second source of error is due to the finite resolution of our simulations
and the finite extraction radius. We estimate both by considering different
extraction radii and performing selected simulations at higher resolution. In
Fig.~\ref{fig:psi4_conv}, we compare the extracted $\Psi_{4}^{20}$ for three
selected cases. The figure shows that there is convergence, but the amount and
properties of noise depend on the resolution and the extraction radius. This is
a well-known and common feature in numerical-relativity
calculations~\cite{Zlochower2012}. Electromagnetic waves have more noise, but
they are always subdominant, so they do not affect the overall error. We obtain
an estimate of the accuracy of our calculation by comparing the amount of total
energy radiated for different resolutions and extraction radii. Combining all
sources of error, we estimate that the error in the energy lost by waves to be
of order \SI{10}{\percent}. On the other hand, the horizon properties are stable and
show excellent degree of convergence (the final properties differ by less than
\SI{0.01}{\percent}). So, we assume that the dominant source of error for
the remnant properties is the intrinsic uncertainty in the initial data.
Therefore, we estimate the error in the identification of the parameters of the
final remnant to be a few percent.

\bibliography{ur_collisions_of_charged_bhs, einsteintoolkit}

\end{document}